\documentclass[12pt]{article}
\usepackage{eurosym}
\usepackage{graphicx}
\usepackage{amsthm}
\usepackage{amsmath}
\usepackage{amssymb}
\usepackage{appendix}
\usepackage{color}
\usepackage{sgame}
\usepackage{verbatim}
\usepackage{caption}
\usepackage{subcaption}
\usepackage{mathtools}
\usepackage{dsfont}
\usepackage{float}
\usepackage[margin=1in]{geometry}
\usepackage{natbib}
\usepackage[utf8]{inputenc}
\usepackage[english]{babel}
\usepackage{lscape}
\usepackage[colorlinks=true,allcolors=blue]{hyperref}
\usepackage{booktabs}
\usepackage{longtable}
\usepackage{threeparttable}
\usepackage{dcolumn}
\usepackage{tikz}
\usepackage{xcolor}
\usepackage{istgame}
\usepackage{multirow}
\usepackage{amsfonts}
\usepackage{adjustbox}
\usepackage{istgame}
\usepackage{bbm}
\usepackage{soul}

\usetikzlibrary{calc}
\usetikzlibrary{calc}
\newtheorem{definition}{Definition}

\newtheorem{proposition}{Proposition}

\DeclareMathOperator*{\argmax}{argmax}
\begin{document}

\title{Virtual Observability in Sequential 
Play\thanks{We thank Seong-Gyu Park for research assistantship; Ala Avoyan, Meng-Jhang Fong, 
David Gill, Thomas R. Palfrey, Stephanie W. Wang, and audiences at the 2025 WEAI annual conference and 
the 2025 North American Conference of the Economic
Science Association for their valuable comments.
This study was approved by the Claremont Graduate University IRB (protocol \#5149).}}
\author{C. Monica Capra\thanks{Claremont Graduate University, Claremont, CA 91711 and Center for the Philosophy of Freedom, University of Arizona, Tucson, AZ 85721.  Email: monica.capra@cgu.edu}
\and 
Charles A. Holt\thanks{%
University of Virginia, Charlottesville, VA 22904. Email: 
charles.holt4@virginia.edu}
\and
Po-Hsuan Lin\thanks{%
University of Virginia, Charlottesville, VA 22904. Email: 
plin@virginia.edu} }
\date{\today}
\maketitle

\begin{abstract}


When players make sequential decisions that are unobservable to one another, their behavior can nonetheless be influenced by knowing who moves first. This sequential structure, often referred to as ``virtual observability,'' suggests that timing alone can shape expectations and choices, even when no information is revealed. The original notion of virtual observability, however, is an equilibrium refinement based on the timing structure and has no bite in games with a unique equilibrium. In this paper, we experimentally examine whether timing still affects behavior in such games, using the Traveler’s Dilemma and the Trust Game.
We find that in the sequential Traveler’s Dilemma without observability, first movers tend to behave closer to the equilibrium prediction than in the simultaneous version. In contrast, timing without observability has no effect on behavior in the Trust Game.


\end{abstract}

\renewcommand{\baselinestretch}{1.2}

\thispagestyle{empty}

\bigskip



\newpage\setcounter{page}{1} \renewcommand{\baselinestretch}{1.2}

\begin{center}
\textit{Simply knowing that others have moved earlier is \\
cognitively similar to having observed what they did.}
\end{center}

\rightline{---Colin F. Camerer (\citeyear{camerer1997progress}), p. 177}

\section{Introduction}

Consider a simple strategic interaction: you must decide whether to cooperate or defect, knowing that another player faces the same choice. If you move first and the other player moves second, but cannot observe your choice, does the timing matter? Standard game theory says ``no,'' since your action is unobservable, the other player's decision cannot depend on your actual choice. Indeed, the sequential game with unobservable actions has the same strategic structure as the simultaneous game, and behavior should be identical.

Yet, the invariance prediction does not match intuition. As a first mover, one might reason: ``Even though the other player cannot see my choice, she will anticipate what I did, and if she anticipates I cooperate, she will cooperate too." This reasoning treats the second mover \emph{as if she could observe} the first mover's action because the timing structure makes such an inference psychologically salient. The first mover behaves as if her choice casts a shadow forward in time, influencing the second mover's expectations and responses.

Colin Camerer introduced the general idea of \emph{Virtual Observability} as: ``...Simply knowing that others have moved earlier is cognitively similar to having observed what they did.'' \cite[p.~177]{camerer1997progress}. In a later study, \cite{weber2004timing} formalized the intuition behind virtual observability as an equilibrium refinement. Their approach proceeds as follows: (i) fix a game of imperfect information in which previous moves are unobservable; (ii) maintaining the temporal order of moves, assume that all previous moves are observable; (iii) find any subgame-perfect equilibria of the game with observable moves; (iv) If any such equilibrium is also a Nash equilibrium of the original game, then select this equilibrium of the original game; (v) otherwise, timing will not affect play. This refinement embodies two essential elements: virtual observable subgame perfection, which assumes that prior moves are virtually observable, and the Game-Nash refinement, which ensures that an equilibrium with virtually observable moves is also a Nash equilibrium of the original game. The latter prevents behavior from being pulled away from the Nash equilibrium.

In this paper, we distinguish the \emph{Virtual Observability Refinement} from a more general perspective that we call \emph{General Virtual Observability}. Under General Virtual Observability, first movers adjust their behavior based on the assumption that prior moves are virtually observable, as in step (iii), \emph{without imposing the Game-Nash Requirement of step (iv)}. The difference between the two concepts can be illustrated in the context of a weak prisoner’s dilemma shown in Figure \ref{fig:magical_example}. In this game, mutual cooperation would yield the highest payoffs for both players, but it is weakly dominated. If player 1 moves first and player 2 moves second without observing 1's actions, standard game theory predicts no change; the equilibrium remains mutual defection. However, player 1 may reason that his move is virtually observable and behave as if player 2 will best respond to his observed action, even though this belief is inconsistent with the Nash equilibrium of the original game. This distinction matters in games with a unique Nash equilibrium like this one: the Virtual Observability Refinement predicts \emph{no timing effects}, since virtual observability is \emph{not} a Nash equilibrium of the original game, whereas General Virtual Observability allows behavior to deviate from the Nash predictions.

In our study, we use laboratory experiments to assess which of these perspectives better explains the behavior of financially motivated participants. We investigate this question using two canonical games: the Traveler's Dilemma \citep{basu1994traveler, capra1999anomalous} and a modified Trust Game \citep{berg1995trust}. Both games feature unique Nash equilibrium, making them ideal for distinguishing between the Virtual Observability Refinement and the General Virtual Observability perspectives on sequential moves with unobservable actions.

In our experiment, we implement two versions of the Traveler's Dilemma: a simultaneous version and a sequential version with unobservable actions. Under General Virtual Observability, a first mover who believes the second mover will undercut should claim the maximum. In the modified Trust Game, an investor and trustee make decisions about how much to invest and return, with transfers occurring only if the return meets a minimum threshold. We implement two sequential versions: investor-first and trustee-first, both with unobservable actions. Under General Virtual Observability, timing should matter asymmetrically: an investor who moves first and believes the trustee will match should invest fully, while a trustee who moves first should return nothing regardless of beliefs about virtual observability.

We employ a within-subject design whereby participants made one-shot decisions in all four games without feedback. In the Traveler's Dilemma, we find significant evidence of timing effects: first movers claim less than second movers 
on average (5.57 vs. 6.14), and first movers choose the equilibrium claim at significantly higher rates (37\% vs. 15\%). This pattern is inconsistent with the Virtual Observability Refinement, which predicts no timing effects. Interestingly, the direction also differs from General Virtual Observability's prediction of maximum claims. Instead of moving away from Nash equilibrium, first movers appear to become more strategically sophisticated, moving toward the Nash equilibrium. 

In contrast, the Trust Game exhibits no significant timing effect. Neither investor nor trustee behavior varies systematically with move order. This null result echoes \citet{weber2004timing}, who found only modest timing effects in ultimatum bargaining, and suggests that virtual observability may be muted in games where other-regarding preferences dominate strategic considerations.

Our findings contribute to the literature in several ways. First, we demonstrate that timing effects can emerge in games with unique Nash equilibria, extending the domain of virtual observability beyond coordination games. Second, we show that neither the Virtual Observability Refinement nor General Virtual Observability fully captures observed behavior—timing matters, but not always in the predicted direction. Third, we document an important boundary condition: timing effects appear stronger in the Traveler's Dilemma than in the Trust Game, suggesting that the cognitive mechanisms underlying virtual observability interact with the motivational structure of strategic environments.

The rest of the paper proceeds as follows. Section \ref{sec:lit_review} reviews the related literature. Section \ref{sec:model} formally introduces our framework of General Virtual Observability. Section \ref{sec:games} describes the Traveler’s Dilemma and a modified Trust Game---the two games studied in our experiment---and outlines the predictions of both the Virtual Observability Refinement and General Virtual Observability. Section \ref{sec:exp_design} presents our experimental design. Section \ref{sec:exp_result} reports the experimental results. Finally, Section \ref{sec:conclusion} concludes. The complete experimental instructions are provided in \ref{sec:instruction}.

\section{Literature Review}\label{sec:lit_review}


The term ``virtual observability'' was first coined by \citet{weber2004timing} as an equilibrium refinement concept, building on \citet{amershi1989manipulated}'s earlier work on manipulated Nash equilibrium---an equilibrium selection process in extensive games where players behave as if previous moves were observable, even though they are actually not. The concept was also motivated by \citet{cooper1993forward}'s influential battle-of-the-sexes experiment, which revealed a striking puzzle: the mere knowledge that one player moved first dramatically affects coordination outcomes even when actions remain unobservable. Comparing simultaneous-move and sequential-move (with unobservable actions) versions of the game, \citet{cooper1993forward} found that simultaneous play converges to a mixed-strategy equilibrium, while sequential play with unobservable actions induces coordination on the first-mover preferred equilibrium.


Since \citet{cooper1993forward}'s study, a growing body of experimental literature has examined the timing effect in various games. \citet{weber2004timing} provide evidence from ``weak link'' coordination games, while \citet{rapoport1997order} documents similar effects in threshold public goods experiments. \citet{budescu1997effects} and \citet{rapoport1998coordination} report findings from a resource dilemma experiment and a three-person coordination game, respectively. Collectively, these studies document a common pattern: under virtual observability, players tend to coordinate on the equilibrium preferred by the earlier mover.

However, these games all feature multiple equilibria where virtual observability can serve as a selection criterion. Beyond this class of games, the findings are mixed. While \citet{guth1998limitations} find no timing effects in dominance-solvable games, \citet{huck2005burning} document a strong timing effect in a control treatment of their burning money experiment.\footnote{Specifically, \citet{huck2005burning} show a timing effect in 
one of their control treatments in which the first mover burns \emph{zero} dollars. In 
this case, the first mover’s decision merely selects which coordination game to play, 
rather than signaling intent in the sense of forward induction. Surprisingly, despite the 
first choice being materially irrelevant, players were significantly more likely to 
coordinate on the first mover’s preferred equilibrium in both subgames (burning the money 
or not).} More recently, \citet{penczynski2016strategic} demonstrates timing effects in a hide-and-seek game with a unique mixed-strategy 
equilibrium.\footnote{In this experiment, players participated in teams of two and were allowed to communicate by submitting a suggested decision along with a message justifying it. Therefore, the observed timing effect also suggests that virtual observability persists even when the game is played in teams.} The author finds that players exhibit higher levels of sophistication in the sequential version with unobservable actions than in the simultaneous version.

The psychological literature offers partial explanations of the mechanisms underlying virtual observability. \citet{camerer1994ambiguity} 
study a sequential matching pennies game with unobservable actions
and find that the ``mismatcher'' prefers a pure strategy (Head or Tail) 
when moving first but prefers to delegates to randomization
when moving second. This finding suggests that individuals 
exhibit different willingness to bet on events that have not yet 
occurred versus events that have already occurred but remain unknown.

\citet{abele2004social} 
propose that timing structure activates different \emph{schemata}. A game structure with simultaneous moves is more likely to activate schemata associated with games of chance or luck, as people tend to
feel they have no control over events that occur simultaneously. In contrast, a sequential-move structure with unobservable actions is more likely to activate schemata related to social 
interaction---namely, that even if earlier actions are unknown, the mere presence of a sequence cues individuals to perceive
interdependence in payoffs. 

\citet{abele2005timing} argue that the \emph{feeling of groupness} can also lead to different behaviors in simultaneous-move games 
and sequential-move games with unobservable actions. Specifically, this hypothesis 
predicts that in a cooperation game, such as a public goods game, 
the simultaneous-move structure is more likely to promote collective thinking 
and foster greater cooperation than the sequential-move structure 
with unobservable actions. Both the schemata hypothesis 
and the feeling of groupness hypothesis find support in experiments
using stag-hunt and public goods games. However, these hypotheses do not 
offer clear predictions beyond the domain of coordination games. 

In summary, previous studies on virtual observability have 
primarily treated it as an equilibrium refinement in sequential 
coordination games with unobserved actions. While there have been 
several attempts to explore the effects of virtual observability in 
other classes of games and to investigate its psychological foundations, 
our understanding remains incomplete. The conflicting findings on the timing effect highlight the need for theoretical frameworks that can predict when and how timing matters across diverse strategic environments. 
To this end, the next section formally defines the notion of \emph{General Virtual Observability}, which extends the original concept beyond an equilibrium selection process.

\section{Framework of General Virtual Observability}\label{sec:model}

In contrast to the Virtual Observability Refinement of \cite{weber2004timing}, our General Virtual Observability is defined as a solution concept for extensive games with unobservable actions,\footnote{Therefore, (General) Virtual Observability has no effect in simultaneous-move games, yielding the same predictions as the standard Nash equilibrium.} which is fundamentally different from standard game-theoretic solution concepts.

For illustration, we define General Virtual Observability within a class of 
two-player extensive games with unobservable actions---specifically, the 
class of games used in our experiment. Let $N = \{1, 2\}$ denote the set 
of players, and let $A_1$ and $A_2$ be the action sets for player 1 and player 2, 
respectively. In this class of games, player 1 moves first by 
choosing an action from $A_1$. Then, player 2 selects an action 
from $A_2$, knowing that player 1 has moved but without observing
which action was taken. In other words, player 2 has a single information set
that includes all actions in $A_1$. The game ends after player 2 makes a decision, 
and the payoff functions for the two players are $u_1$ and $u_2$, respectively. 
Figure \ref{fig:generic_tree} below presents a generic game tree for this class of 
games.

\begin{figure}[htbp!]
    \centering
    \includegraphics[width=0.6\linewidth]{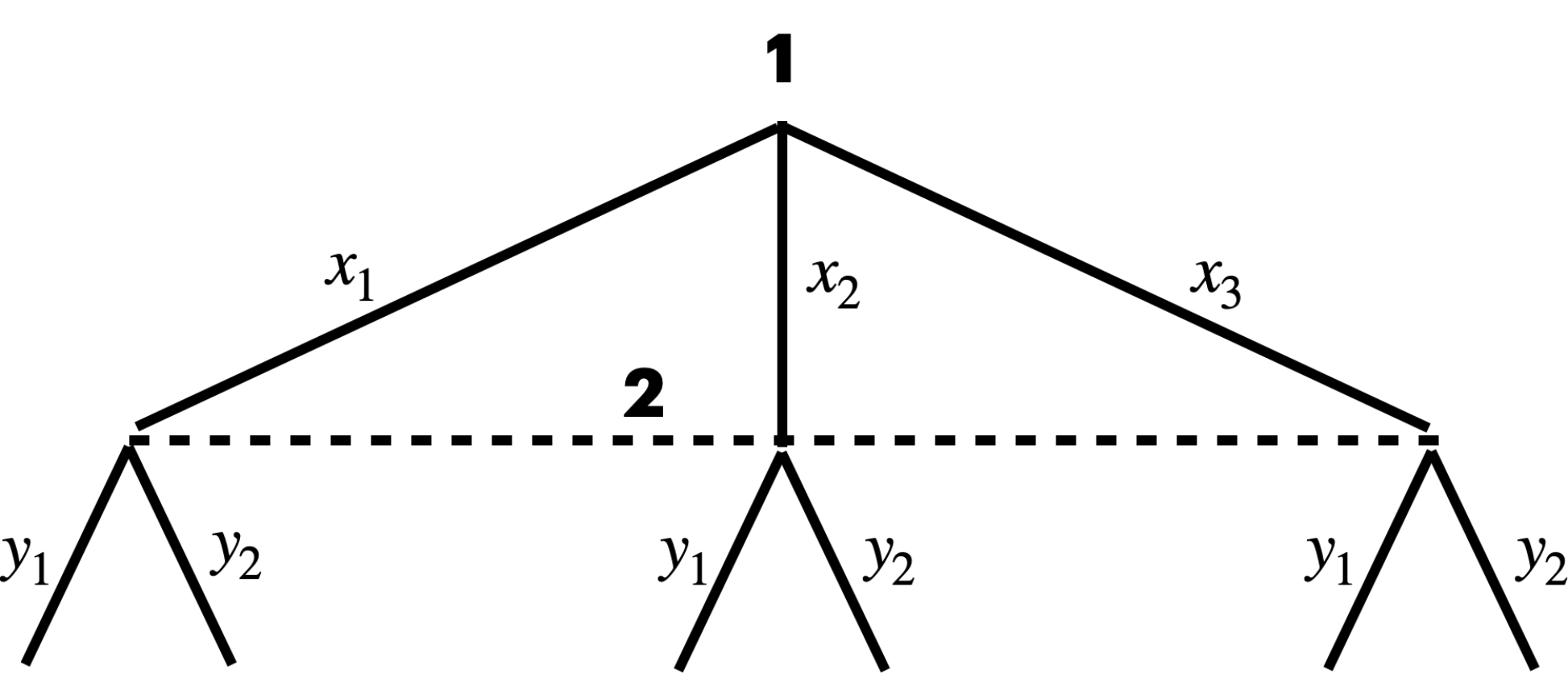}
    \caption{A generic game tree for a class of 
    two-player extensive games with unobservable actions, 
    where $A_1 = \{x_1, x_2. x_3\}$ and $A_2 = \{y_1, y_2\}$.}
    \label{fig:generic_tree}
\end{figure}

General Virtual Observability is defined as an 
assessment together with a \emph{virtual conjecture} 
made by player 1 about player 2's behavioral strategy. 
The assessment consists of a behavioral strategy 
profile $(\sigma_1, \sigma_2)$, where $\sigma_i \in \Delta(A_i)$ 
for $i \in N$, and a belief $\mu \in \Delta(A_1)$ held
by player 2 about player 1's action.
Moreover, under General Virtual Observability, player 1 mistakenly believes 
that player 2 can observe player 1’s unobservable action. 
Formally, player 1 incorrectly believes that player 2’s 
behavioral strategy is measurable with respect to histories 
rather than information sets.
That is, player 1’s virtual conjecture about player 2’s 
behavioral strategy is a function 
$\tilde{\sigma}_2 \colon A_1 \rightarrow \Delta(A_2)$.

In addition, General Virtual Observability imposes 
consistency conditions on player 1's virtual conjecture 
and player 2's belief. Players under General Virtual Observability are (sequentially) rational. Accordingly, player 1 incorrectly believes
that player 2 best responds at every history, and best responds to this virtual conjecture. In other words, player 1 behaves as if he \emph{erases} the information set and performs backward induction in the corresponding extensive game with perfect information.
Furthermore, player 2 is assumed to have a correct perception of player 1's behavioral strategy and forms her belief using Bayes' rule. Consequently, under General Virtual Observability, player 2's belief is $\mu = \sigma_1$, and she best responds to this belief.
Definition \ref{def:magical_eq} formally defines our General Virtual Observability.

\begin{definition}\label{def:magical_eq}
General Virtual Observability in a two-player extensive game with unobservable actions is an assessment and player 1's virtual conjecture, 
$\langle  (\sigma^*_i)_{i\in N}, \mu; \; \tilde{\sigma}_2\rangle$ where,
\begin{itemize}
    \item[(i)]  For any $a_1\in A_1$ and any $a_2' \in A_2$ such that 
    $\tilde{\sigma}_2(a_2'|a_1) > 0$, 
    $a_2' \in \argmax_{a_2\in A_2} u_2(a_1, a_2);$
    \item[(ii)]  For any $a_1' \in A_1$ such that $\sigma_1^*(a_1)>0$,
    $$a_1'\in\argmax_{a_1\in A_1} 
    \sum_{a_2\in A_2} \tilde{\sigma}_2(a_2|a_1) u_1(a_1, a_2);$$
    \item[(iii)] For any $a_1\in A_1$, $\mu(a_1) = \sigma_1^*(a_1)$; and 
    \item[(iv)] For any $a'_2 \in A_2$ such that $\sigma_2^*(a'_2)>0$, 
    $a_2' \in \argmax_{a_2\in A_2} \sum_{a_1\in A_1} \mu(a_1)u_2(a_1, a_2)$.
\end{itemize}

\end{definition}

From the definition of General Virtual Observability, we can see that for games with multiple equilibria, General Virtual Observability reduces to the original notion of virtual observability introduced by \citet{amershi1989manipulated} and \citet{weber2004timing}---selecting the Nash equilibrium that remains an equilibrium after erasing all information sets. Consequently, if a game has a unique Nash equilibrium, the original virtual observability refinement has no bite, predicting that the timing of moves does not affect play. Yet General Virtual Observability may still diverge from the standard Nash equilibrium in this case.

\begin{figure}[htbp!]
    \centering
    \includegraphics[width=\linewidth]{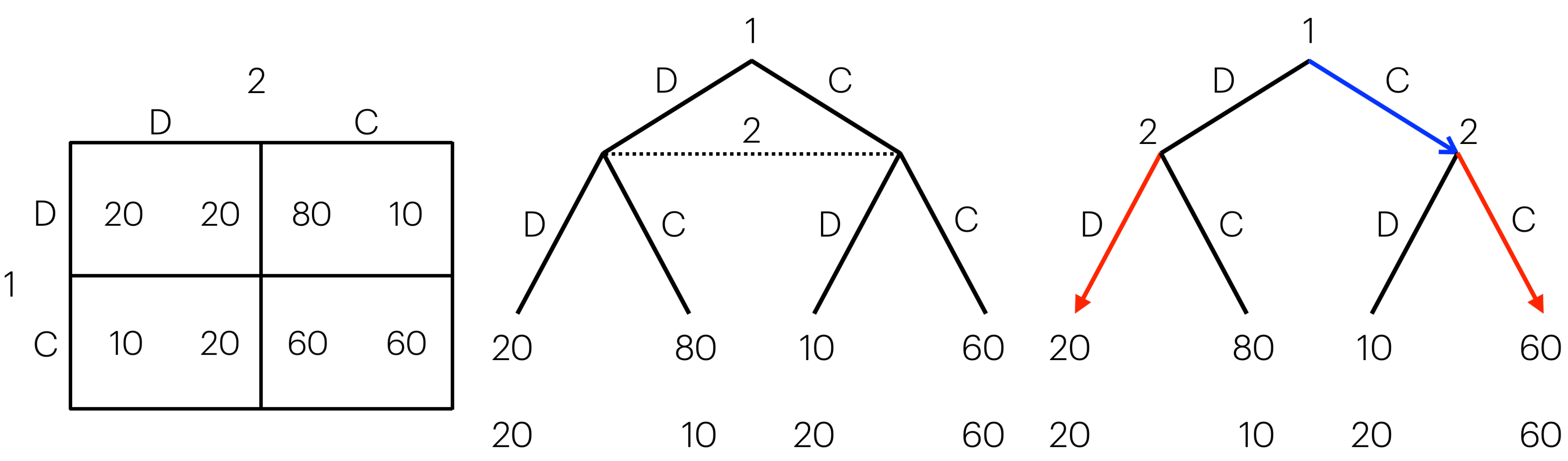}
    \caption{(Left) Normal form of the weak prisoner’s dilemma. 
    (Middle) Extensive form of the sequential version. 
    (Right) Red arrows indicate player 1’s virtual conjecture about player 2; 
    the blue arrow shows player 1’s best response.} 
    \label{fig:magical_example}
\end{figure}

Specifically, General Virtual Observability is defined as a fixed point in which players hold incorrect perceptions about others’ future actions. More importantly, all players under General Virtual Observability understand how others form their virtual conjectures. This distinction makes General Virtual Observability a solution concept that is fundamentally different from Nash equilibrium.\footnote{Our notion of General Virtual Observability is adapted from the ``Magical Equilibrium'' proposed by \citet{lin2025magical}, an extensive-form solution concept that captures the 
\emph{illusion of causality}---the cognitive bias of mistakenly perceiving a causal link between unrelated events. In addition, it is worth noting that General Virtual Observability is also related to the ``Magical Thinking'' framework of \citet{daley2017magical}, as their model likewise posits that players behave as if their opponents' behavior varies with their own unobservable choices. However, they assume that players incorrectly believe their opponents will \emph{copy} their unobservable actions, rather than best respond, which distinguishes General Virtual Observability from their magical thinking model.}
To illustrate how General Virtual Observability differs from Nash equilibrium, consider a sequential weak prisoner’s dilemma game, with its normal and extensive forms shown in Figure \ref{fig:magical_example}. The game has a unique Nash equilibrium in which both players defect; therefore, both virtual observability and Nash equilibrium predict that the timing of moves does not affect play in this game.

General Virtual Observability, on the other hand, predicts that if player 1 moves first, both players will choose to cooperate. The intuition is that player 2 has an incentive to ``copy'' player 1’s action. Consequently, player 1's virtual conjecture is that player 2 will mirror his move, making cooperation the best response for player 1. Given that player 1 chooses to cooperate, player 2’s best response is also to cooperate, thus generating the prediction of General Virtual Observability.

This example illustrates that General Virtual Observability generally differs from both the Nash equilibrium and the Virtual Observability Refinement. 
As such, it provides a useful tool for studying the effects of move timing even in games with a unique Nash equilibrium. To this end, we adopt two games with a unique Nash equilibrium in our experiment---the Traveler’s Dilemma and a modified Trust Game---which provide a clear benchmark for the Nash equilibrium.

\section{Games and Hypotheses}  \label{sec:games}

Our study investigates the effect of virtual observability in two canonical experimental games: the Traveler’s Dilemma and a modified Trust Game. Both games have a unique Nash equilibrium, yet prior studies have shown that behavior in these games is often inconsistent with standard game-theoretic predictions.

\subsection{Traveler's Dilemma}

The original Traveler's Dilemma was 
introduced by \cite{basu1994traveler} as a parable to 
highlight how standard game-theoretic reasoning can lead to highly 
unintuitive outcomes. It was later studied experimentally by 
\cite{capra1999anomalous} and \cite{goeree2001ten}, who demonstrated 
that observed behavior is inconsistent with the standard Nash 
equilibrium prediction.

The story behind the dilemma is as follows: Two travelers purchase identical antiques 
while on a tropical vacation. During the return trip, both antiques are damaged, and 
the airline asks each traveler to independently report the value of the antique. Let the two travelers be denoted 
by $i$ and $j$, and let $n_i$ and $n_j$ be their reported values, respectively. If both travelers report the same value ($n_i = n_j$), it is 
reasonable to assume that they are telling the truth, and the airline
compensates each of them with that reported amount. 
However, if traveler $i$
reports a number strictly higher than traveler $j$ ($n_i > n_j$), 
it is reasonable for the 
airline manager to assume that traveler $j$ is honest and traveler $i$ is lying. 
In that case, the airline treats the lower number $n_j$ as the true 
value of the antique and pays traveler $j$ the amount $n_j$ plus a 
reward. In contrast, traveler $i$ receives $n_j$ minus a 
penalty for lying about the value.

In our implementation of the Traveler's Dilemma, players 
may claim any
amount between 4 and 8 units in increments of 0.5, and the reward/penalty
is set to 1 unit. The unique Nash equilibrium is for both travelers to
choose the lowest number, 4.
The intuition is that, irrespective of the antique’s actual value, each
player has a unilateral incentive to ``undercut'' the other’s claim.
Consider the case where $n_i > n_j$, so traveler $i$ receives a payoff
of $n_j - 1$. In this case, traveler $i$ can profitably deviate to
$n_j - 0.5$, which yields a higher payoff of $n_j + 0.5$. This
undercutting logic applies whenever a player’s claim exceeds the
minimum, leading both travelers to choose the lowest number, the
unique equilibrium outcome.\footnote{\cite{basu1994traveler} noted that
this is also the unique rationalizable equilibrium.}

\begin{figure}[htbp!]
    \centering
    \includegraphics[width=0.6\linewidth]{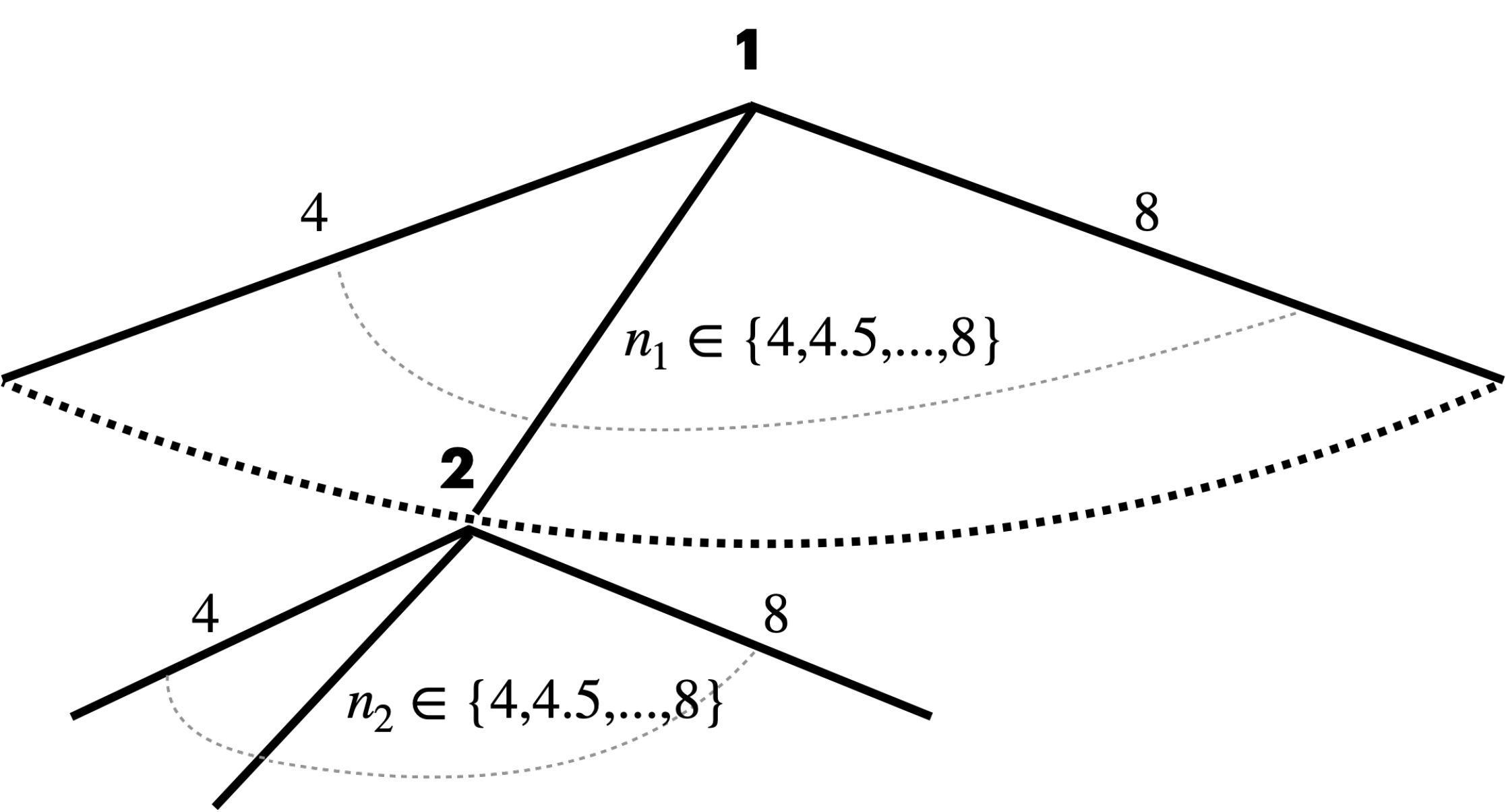}
    \caption{The game tree of the sequential Traveler's Dilemma}
    \label{fig:seq_td_tree}
\end{figure}

To study the effect of virtual observability, we consider two 
versions of the Traveler's Dilemma in our experiment: 
A simultaneous-move version and a sequential version with unobservable actions. 
In contrast to the simultaneous-move game, in the sequential game, 
Traveler 1 makes a claim first, and Traveler 2 then chooses their claim 
\emph{without knowing} the amount claimed by Traveler 1. Thus, the sequential Traveler's Dilemma is effectively an extensive game with 
imperfect information, as illustrated in Figure \ref{fig:seq_td_tree}.




Since the information available to players at the time of their decisions is the same 
in both the simultaneous and sequential versions, the Nash equilibrium is identical 
across the two. Following the notion of virtual observability in \cite{weber2004timing},
the timing structure should have no impact on behavior. 

\bigskip

\noindent\textbf{Hypothesis 1} (Virtual Observability in Traveler’s Dilemma).
\emph{\begin{itemize}
    \item[(i)] In the sequential game, the first movers’ claims and the second movers’ claims are statistically indistinguishable.
    \item[(ii)] Both first and second movers’ claims are statistically indistinguishable from those in the simultaneous game.
\end{itemize}}

In contrast, General Virtual Observability predicts that players’ behavior will strikingly differ between the two versions. In the sequential version, Traveler 1 will choose 8 and Traveler 2 will choose 7.5, whereas in the simultaneous version, both players will choose 4, since (General) Virtual Observability has no effect in simultaneous-move games.  

The intuition behind General Virtual Observability in the sequential Traveler’s Dilemma is that, under virtual observability, Traveler 1 mistakenly believes that Traveler 2 will undercut regardless of his choice. As a result, Traveler 1 perceives the penalty as unavoidable and chooses 8 to maximize the value of the antique. Given Traveler 1’s strategy, Traveler 2’s best response is to choose 7.5. The formal statement of the result and its proof are provided in  \ref{sec:proof}. Moreover, this theoretical prediction leads to the following hypothesis. 

\bigskip

\noindent\textbf{Hypothesis 1$'$} (General Virtual Observability in Traveler’s Dilemma).
\emph{\begin{itemize}
    \item[(i)] In the sequential game, first movers make larger claims than second movers.
    \item[(ii)] Both first and second movers make larger claims in the sequential game than in the simultaneous game.
\end{itemize}}

\subsection{A Modified Trust Game}

Since the introduction of the ``trust game'' by \cite{berg1995trust}, it has 
become the standard laboratory experiment for measuring trust.
In the canonical trust game, an \emph{investor} is randomly and 
anonymously paired with a \emph{trustee}. Both the investor and the 
trustee receive an endowment $E>0$. At the beginning, the investor may send 
some or all of their endowment to the trustee (the second mover). 
The amount sent is then tripled by the experimenter. After observing the amount 
received, the trustee can choose to return some or all of it to the investor.

The unique subgame-perfect Nash equilibrium of the canonical trust game 
is for the trustee to return nothing, regardless of the amount invested 
by the investor; anticipating this, the investor chooses to invest nothing. 
Relative to this equilibrium benchmark, positive investments from the investor are 
interpreted as a measure of trust, and returns from the trustee as a measure of 
trustworthiness.

The interpretation of trust and trustworthiness is, in fact, independent 
of the timing of the game. That is, even if the trust game 
is played simultaneously in normal form, players would make
the same \emph{hypothetical} inference as if the game were played sequentially in 
extensive form. We leverage this insight to study the interaction
between virtual observability and both trust and trustworthiness.
Specifically, we consider a modified trust game in which the investor and trustee decide how much to invest and how much to return, respectively, but decisions are unobserved.  The game is represented in normal form in 
Table \ref{tab:payoff_trust}. In this game, both the investor and the trustee
have an endowment of $E=4$ units. The investor may 
invest any integer amount from 0 to 4, denoted by 
$I\in\{0,1,2,3,4\}$. At the same time, the trustee decides how 
much to return to the investor, choosing an integer from 0 to 4, denoted by 
$R\in \{0,1,2,3,4 \}$. We implement a minimum return requirement. The transfer occurs only if the trustee’s return is \emph{no less than} 
the investor’s investment, i.e., $R\geq I$;
otherwise, both players receive only half of their initial endowments. 
If a transfer occurs, 
the investor’s payoff is $E - I + 2R$ and 
the trustee’s payoff is $E + 3I - 2R$.\footnote{To mitigate potential
confusion about the payoff structure, the trustee's action set in the
experiment is restricted to multiples of 2, i.e., $\{0, 2, 4, 6, 8\}$.
See \ref{sec:instruction} for the experimental instructions.}

\begin{table}[htbp!]
\centering
\caption{The normal form of our modified trust game}
\label{tab:payoff_trust}
\renewcommand{\arraystretch}{1.15}
\begin{tabular}{ccccccc}
 &  & \multicolumn{5}{c}{Trustee $R$} \\ \cline{2-7} 
\multicolumn{1}{c|}{} & \multicolumn{1}{c|}{} & 4 & 3 & 2 & 1 & \multicolumn{1}{c|}{0} \\ \cline{2-7} 
\multicolumn{1}{c|}{} & \multicolumn{1}{c|}{4} & $8, 8$ & $2, 2$ & $2, 2$ & $2, 2$ & \multicolumn{1}{c|}{$2, 2$} \\
\multicolumn{1}{c|}{} & \multicolumn{1}{c|}{3} & $9, 5$ & $7, 7$ & $2, 2$ & $2, 2$ & \multicolumn{1}{c|}{$2, 2$} \\
\multicolumn{1}{c|}{Investor $I$} & \multicolumn{1}{c|}{2} & $10, 2$ & $8, 4$ & $6, 6$ & $2, 2$ & \multicolumn{1}{c|}{$2, 2$} \\
\multicolumn{1}{c|}{} & \multicolumn{1}{c|}{1} & $11, -1$ & $9, 1$ & $7, 3$ & $5, 5$ & \multicolumn{1}{c|}{$2, 2$} \\
\multicolumn{1}{c|}{} & \multicolumn{1}{c|}{0} & $12, -4$ & $10, -2$ & $8, 0$ & $6, 2$ & \multicolumn{1}{c|}{$4, 4$} \\ \cline{2-7} 
\end{tabular}
\end{table}

Based on this normal form, we examine two 
versions of the sequential trust game with unobservable actions. In the first version,
the investor decides how much to invest, and then the trustee chooses 
how much to return, knowing that the investor has already made an unobservable 
investment decision. In the second version, the trustee makes 
the return decision first, and the investor then decides how much to invest, knowing 
that the trustee has already made an unobservable choice.
Despite this difference in timing, the Nash equilibrium remains the same 
across both versions, predicting that both the investor and the trustee choose 0. 
As a result, the notion of virtual observability in \cite{weber2004timing} again
yields a null prediction: manipulating the timing structure should have no impact on
behavior, leading to the following hypothesis. 

\bigskip

\noindent\textbf{Hypothesis 2} (Virtual Observability in Trust Game).
\emph{\begin{itemize}
    \item[(i)] The distribution of investment when the investor is the first mover is statistically indistinguishable from the distribution when the investor is the second mover.
    \item[(ii)] The distribution of return when the trustee is the first mover is statistically indistinguishable from the distribution when the trustee is the second mover.
\end{itemize}}

In contrast, General Virtual Observability predicts that in the 
\emph{investor-first} version, both players will choose 4 units, 
whereas in the \emph{trustee-first} version, both will choose 0.
The intuition behind this result is that when the investor moves first, he mistakenly believes
that the trustee can observe the investment and will match the return
accordingly, making it optimal for the investor to invest 4 units 
and for the trustee to return 4 units. In contrast, when the trustee 
moves first, even if she incorrectly believes the investor 
can observe her return decision, her optimal choice is still to return 0, since the 
investor’s best response, regardless of the return, is to invest 0. 
The formal statement and proof of this result are provided in \ref{sec:proof}. 
Based on this prediction, we derive the following testable hypothesis.

\bigskip

\noindent\textbf{Hypothesis 2$'$} (General Virtual Observability in Trust Game).
\emph{\begin{itemize}
    \item[(i)] Investment is higher in the investor-first version than in the trustee-first version.
    \item[(ii)] Return is higher in the investor-first version than in the trustee-first version.
\end{itemize}}

\section{Experimental Design and Implementation}
\label{sec:exp_design}

Our experiment consists of four games: a simultaneous Traveler’s 
Dilemma (TDsim), a sequential Traveler’s Dilemma (TDseq), 
and two versions of the modified sequential Trust Game that differ in which player moves first—the investor-first (IT) and the trustee-first (TI) versions, as described in Section \ref{sec:games}.

To enhance the statistical power for detecting the effect of virtual observability, we employ a \emph{within-subject design} in which each participant plays all four games \emph{once}, alternating between a Traveler’s Dilemma and a Trust Game. To balance any order effects, in half of the sessions participants begin with TDseq followed immediately by the IT version of the Trust Game, while in the other half they begin with TDsim followed immediately by the TI version of the Trust Game.
Therefore, roughly half of the participants
face the sequence TDseq-IT-TDsim-TI, and the other half face
the sequence TDsim-TI-TDseq-IT.
Finally, to avoid potential spillover effects, participants do not receive any feedback between games.

At the beginning of the sequential Traveler’s Dilemma and the two modified Trust Games, each participant is randomly assigned a 
role---first or second mover in the Traveler’s Dilemma, and investor or trustee in the Trust Games.\footnote{In our instructions, we label the investor and the trustee as the “passer” and the “receiver,” respectively.} In each game, participants are randomly matched with an opponent and receive full feedback about others' decisions only at the end of the experiment. The payoffs from one Traveler’s Dilemma and one Trust Game are then randomly selected to determine each participant’s earnings for the session. See \ref{sec:instruction} for the experimental instructions.

In total, we conducted 16 sessions with 108 participants between September and November 2025 at the Claremont Colleges in Claremont, California. The number of sessions and participants for each 
sequence of treatments are summarized in Table \ref{tab:exp_summary}.
The participant sample consists of 44\% males and 44\% economics majors, with an average cash compensation of \$20.86 paid at the end of the experiment. The experiment lasted about 30 minutes.


\begin{table}[htbp!]
\centering
\caption{Summary of numbers of sessions and participants}
\label{tab:exp_summary}
\renewcommand{\arraystretch}{1.3}
\begin{tabular}{cccc}
\hline
Sequence of Treatments &  & \# of Sessions & \# of Participants \\ \hline
TDseq - IT - TDsim - TI &  & 8 & 58 \\
TDsim - TI - TDseq - IT &  & 8 & 50 \\ \hline
Total &  & 16 & 108 \\ \hline
\end{tabular}
\end{table}

\section{Experimental Results}
\label{sec:exp_result}

In this section, we present the experimental results for the Traveler’s Dilemma (Section \ref{subsec:td_result}) and the Trust Game (Section \ref{subsec:trust_result}). When reporting the results, we first present the aggregate distributions of choices and then test the effect of the timing structure by leveraging our within-subject design.

\subsection{Traveler's Dilemma}
\label{subsec:td_result}

\begin{figure}[htbp!]
    \centering
    \includegraphics[width=\linewidth]{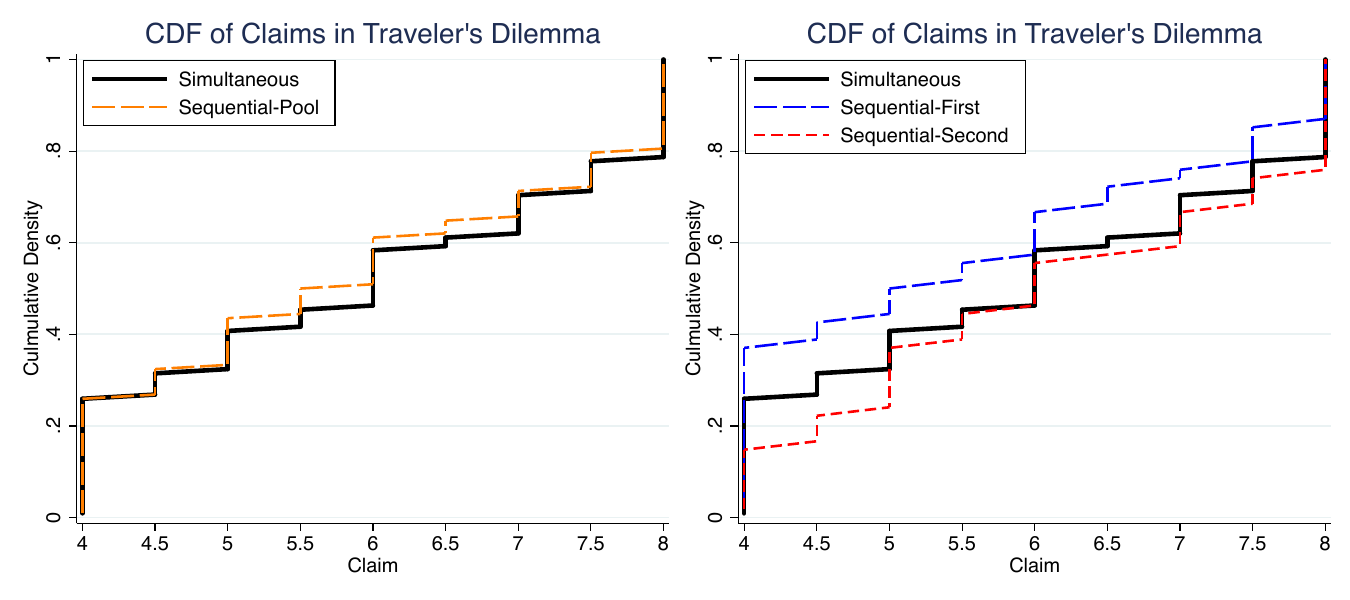}
    \caption{The CDF of claims in the Traveler’s Dilemma. The left panel plots the distribution of claims in the simultaneous treatment and in the sequential treatment (pooling the claims of first and second movers). The right panel separates the distributions of claims for first and second movers in the sequential treatment.}
    \label{fig:td_claim_cdf}
\end{figure}

The left panel of Figure \ref{fig:td_claim_cdf} plots the aggregate distribution of claims in the simultaneous (solid line) and sequential (dashed line) treatments, where we pool the claims of the first and second movers. The average claims in the sequential and simultaneous treatments are 5.86 and 5.94, respectively. Although the average claims in both treatments differ significantly from the unique Nash equilibrium 
(t-test p-value $< 0.001$ in both treatments), they are not significantly different from each other (Rank-sum test p-value $= 0.706$).

While the aggregate distributions of claims do not differ significantly across treatments, we next examine whether the timing structure has different effects on the first and second movers in the sequential treatment. To this end, in the right panel of Figure \ref{fig:td_claim_cdf}, we separate the distributions of claims for the first and second movers. We find that the distribution of the second mover \emph{first-order stochastically dominates} that of the first mover, suggesting that first movers tend to choose smaller claims than second movers (the means for first and second movers are 5.57 and 6.14, respectively; Rank-sum test p-value $= 0.039$).

Additionally, when comparing the frequency of the equilibrium claim between first and second movers, we find that 37\% of first movers choose to claim 4, whereas only 15\% of second movers do so (Rank-sum test p-value $= 0.009$). Moreover, relative to the simultaneous treatment, first movers choose the equilibrium claim significantly more often (26\% in the simultaneous treatment; Signed-rank 
test p-value $= 0.083$), while second movers choose it significantly less often (Signed-rank test p-value $= 0.058$). These observations suggest that in the sequential treatment, first movers not only choose lower claims on average but are also more likely to choose the equilibrium claim than second movers.

Lastly, in our experiment, each subject participates in both 
the sequential and simultaneous treatments, allowing us to 
analyze the timing effect not only at the aggregate level but 
also at the individual level. To analyze this effect at the 
individual level, we define $\Delta^{\mbox{Claim}}_i$ as the 
difference between subject $i$'s
claim in the sequential treatment and their claim in the simultaneous treatment.

\begin{figure}[htbp!]
    \centering
    \includegraphics[width=0.6\linewidth]{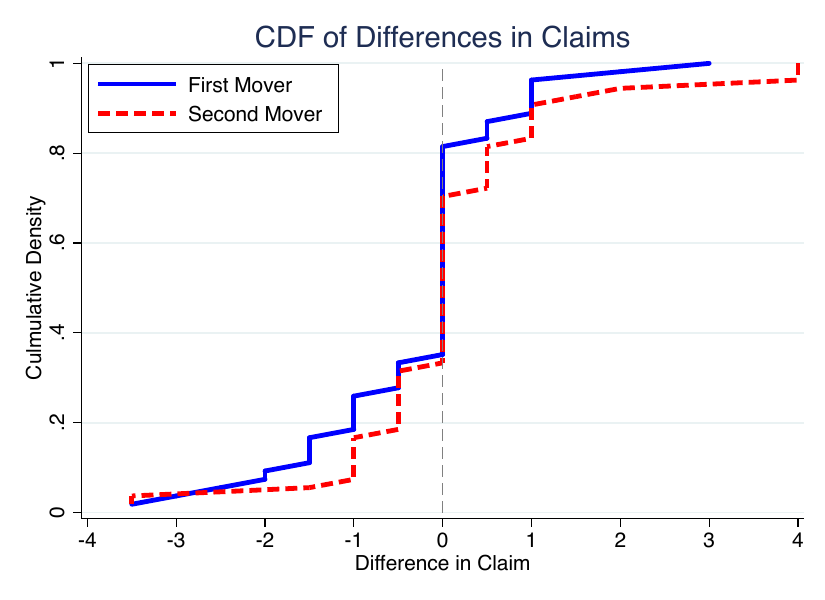}
    \caption{The CDF of the differences in claims between the sequential and simultaneous treatments for the first mover (solid) and the second mover (dashed).}
    \label{fig:td_diff_cdf}
\end{figure}

Figure \ref{fig:td_diff_cdf} plots the distribution of $\Delta^{\mbox{Claim}}_i$ for the first mover (solid line) and the second mover (dashed line). For the second movers, the distribution
of differences is symmetric around zero---31\% of subjects have $\Delta^{\mbox{Claim}}_i < 0$ and 30\% have 
$\Delta^{\mbox{Claim}}_i > 0$---resulting in a mean difference of 0.09, which is not significantly different from 0 (t-test p-value $= 0.614$).
In contrast, 33\% of first movers choose strictly smaller claims in the sequential treatment than in the simultaneous treatment, whereas only 19\% choose strictly greater claims in the sequential treatment. This results in a mean difference of $-0.269$, which is significantly different from zero (one-tailed t-test p-value $= 0.041$).

In summary, we find that in the Traveler’s Dilemma, 
players tend to make smaller claims when they choose before 
the other player, even when they know the other player cannot 
observe their claim.
In contrast, for players who make their claims after the other player, their behavior does not differ significantly from the situation in which both players make their claims simultaneously. In other words, our findings suggest that virtual observation is \emph{more} than an equilibrium selection device. In the Traveler’s Dilemma, virtual observability appears to make first movers behave more sophisticatedly, as their claims are closer to the Nash equilibrium prediction.

The Traveler’s Dilemma is a dominance-solvable game that allows 
us to examine how the timing structure affects players’ revealed 
levels of sophistication. In the following section, we present the 
experimental results for the Trust Game, a fundamentally different 
type of game, as it is not dominance-solvable and 
other-regarding preferences play an important role in shaping 
players’ behavior.

\subsection{Trust Game}
\label{subsec:trust_result}

\begin{figure}[htbp!]
    \centering
    \includegraphics[width=0.9\linewidth]{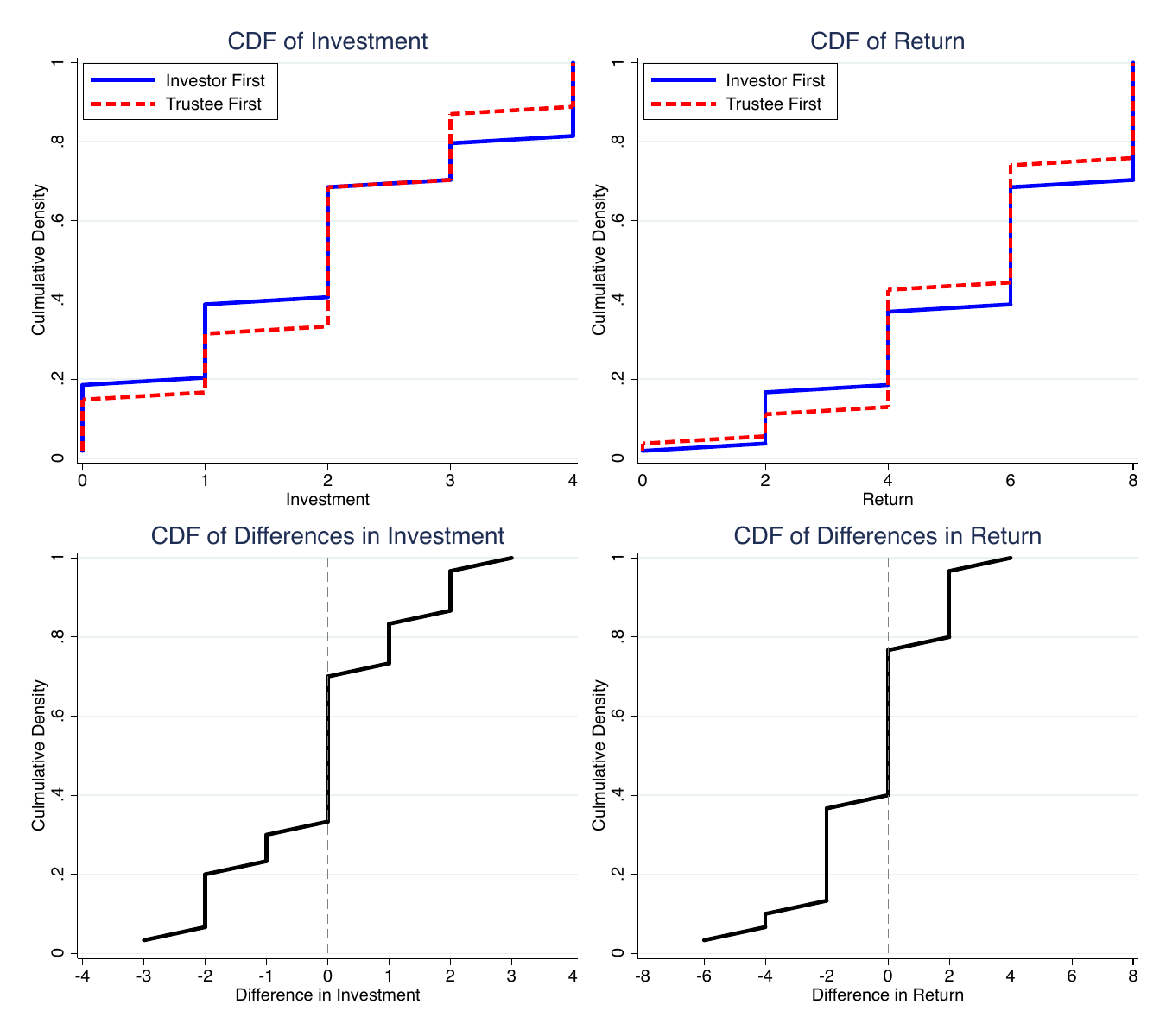}
    \caption{(Top row) The CDF of investment and return in the Trust Game. (Bottom left) The difference in investment between the investor-first and trustee-first treatments. (Bottom right) The difference in return between the trustee-first and investor-first treatments.}
    \label{fig:trust_full_analysis}
\end{figure}

The top two panels of Figure \ref{fig:trust_full_analysis} show the aggregate distributions of investment and return in the investor-first treatment (solid line) and the trustee-first treatment (dashed line). In sharp contrast to the results from the Traveler’s Dilemma, we observe that neither the distribution of investment nor the distribution of return exhibits a first-order stochastic dominance relationship between the two timing structures in the Trust Game.

The average amounts of investment
are 1.94 in the investor-first treatment and 1.98 in the trustee-first treatment, which are not significantly different from each other 
(Rank-sum test p-value $= 0.795$), as investors, on average, make equal-split offers under both timing structures.
Similarly, the mean returns 
do not differ significantly across the two timing structures, 
with means of 5.52 in the investor-first treatment and 5.37
in the trustee-first treatment (Rank-sum test p-value $= 0.660$), 
both of which are slightly lower than three times the 
equal-split offer.\footnote{The average amount of investment is consistent with the 
canonical findings of \cite{berg1995trust}, where investors typically invest about half
of their endowment. However, the average return in our experiment is higher than 
what is commonly reported in the literature. We hypothesize that this difference
is due to our minimum return requirement.}

From the aggregated distributions, we find that the behavior of 
both investors and trustees is largely insensitive to the manipulation of the timing structure of the game. To examine the robustness of this null result, we further analyze the data at the individual level. Specifically, we define $\Delta^{\mbox{Invest}}_i$ as the difference between investor 
$i$'s investment in the investor-first treatment and their investment in the trustee-first treatment. Similarly, we define 
$\Delta^{\mbox{Return}}_i$ as the difference between trustee $i$'s return in the trustee-first treatment and their return in the 
investor-first treatment.\footnote{This individual-level analysis focuses on the 60 subjects who retained the same role (either investor or trustee) in both versions of the Trust Game.}

The bottom two panels of Figure \ref{fig:trust_full_analysis} 
plot the distributions of $\Delta^{\mbox{Invest}}_i$ and $\Delta^{\mbox{Return}}_i$. From these figures, 
we see that both distributions are symmetric around zero, resulting in mean values of $-0.033$ for $\Delta^{\mbox{Invest}}_i$ and $-0.467$ for $\Delta^{\mbox{Return}}_i$, neither of which is significantly different from zero (t-test p-value $= 0.901$ for $\Delta^{\mbox{Invest}}_i$; t-test 
p-value $= 0.243$ for $\Delta^{\mbox{Return}}_i$).
Based on these observations, we conclude that neither the 
investor’s nor the trustee’s behavior varies systematically
depending on whether they move first or second.

Our finding of a null timing effect in the Trust Game echoes the results of \cite{weber2004timing}, who documented only a modest timing effect in the ultimatum game. In their experiment, they consider three versions of the game: proposer-first, responder-first, and simultaneous,\footnote{Similar to our Trust Game experiment, \cite{weber2004timing} implemented the ultimatum game using the strategy method, in which the responder is asked to choose a minimum acceptable offer (MAO).} finding that although the timing structure has a small effect on responders’ behavior, the distribution of proposers’ offers is nearly identical across all versions. \cite{weber2004timing} hypothesize that the timing effect is substantially weaker because of strong fairness preferences. In this sense, our result complements \cite{weber2004timing}, confirming that in games where trust and trustworthiness play an important role, the timing effect is almost ``muted.''

\section{Conclusion}
\label{sec:conclusion}

Our experiments investigate whether the timing of unobserved moves affects strategic behavior in games with unique Nash equilibria. The Virtual Observability Refinement of \cite{weber2004timing} imposes the Game-Nash Requirement: any equilibrium under virtual observability must also be a Nash equilibrium of the original game. More generally, when a game has a unique Nash equilibrium, this refinement has no bite, predicting that timing should not affect play. By contrast, General Virtual Observability allows for timing effects even when the equilibrium is unique. Indeed, first movers may behave as if their choices cast a shadow forward in time, influencing the second mover’s expectations and responses.

Our findings reject the prediction of the Virtual Observability Refinement in the Traveler's Dilemma but not in the modified Trust Game. 

In the Traveler’s Dilemma, timing affects behavior. In the sequential treatment, first movers claim significantly less on average than second movers, and they choose the Nash equilibrium claim at more than twice the rate of second movers (37\% vs. 15\%). Relative to the simultaneous treatment, in which 26\% of participants choose the minimum claim, first movers in the sequential treatment claim 4 significantly more often, whereas second movers choose it significantly less often. Our within-subject design allows us to examine individual-level responses to timing: 33\% of first movers claim strictly less in the sequential game than in the simultaneous game, while only 19\% claim strictly more. The mean within-subject difference for first movers is significantly different from zero. For second movers, the distribution of differences is symmetric around zero and not significantly different from zero.

Although we observe a timing effect in the Traveler’s Dilemma, it does not occur in the expected direction. General Virtual Observability predicts higher claims; however, we find that first movers move closer to the Nash equilibrium rather than toward the prediction of General Virtual Observability. This result suggests that the psychological mechanism underlying the timing effect may involve heightened strategic scrutiny rather than a specific belief that one’s action will be observed and best-responded to. The pattern we observe---first movers becoming more strategically sophisticated under sequential timing---is consistent with \citet{penczynski2016strategic}, who found that first movers exhibit higher levels of strategic thinking in sequential versions of a hide-and-seek game compared to simultaneous versions.

In the Trust Game, we find no evidence of timing effects. Average investment is nearly identical whether the investor moves first (1.94) or second (1.98), and average returns are similarly invariant across timing conditions (5.52 in investor-first vs. 5.37 in trustee-first). Individual-level analysis confirms this null result: the distributions of within-subject differences are symmetric around zero for both roles. Thus, neither aggregate nor individual behavior responds to the manipulation of move order in the modified Trust Game.

This null result is consistent with the Virtual Observability Refinement, but it is also compatible with General Virtual Observability being overwhelmed by other forces. Indeed, our Trust Game results echo \citet{weber2004timing}, who documented only modest timing effects in ultimatum bargaining, where fairness norms are also salient. They hypothesize that strong fairness preferences may ``crowd out'' virtual observability effects. Similarly, \citet{guth1998limitations} found no timing effects in matrix games with unique equilibria when dominance was weak. Together with our findings, these results suggest a boundary condition: virtual observability may be muted in games where other-regarding preferences or weak strategic incentives reduce the salience of move order.

Three broader lessons emerge from our experiments. First, the domain of virtual observability extends beyond coordination games: timing can affect behavior even when the equilibrium is unique, contrary to the predictions of the Virtual Observability Refinement. This expands the empirical scope of the phenomenon documented by \citet{cooper1993forward} and subsequent work in coordination settings. Second, the direction of timing effects need not follow the predictions of General Virtual Observability. In the Traveler’s Dilemma, first movers become more sophisticated, not less---they move toward the Nash equilibrium rather than toward the high claims predicted by backward induction under virtual observability. This finding suggests that virtual observability operates through strategic activation rather than the specific belief structure formalized in step (iii) of \citet{weber2004timing}. Third, timing effects appear sensitive to the motivational structure of the game. When social preferences are central, as in the Trust Game, the psychological salience of move order may be overshadowed by concerns about fairness and reciprocity, consistent with the crowding-out hypothesis of \citet{weber2004timing}.

Our findings raise several questions for future research. What features of strategic structure determine when timing effects emerge and in which direction they operate? Can the sophistication-enhancing effect we observe in the Traveler’s Dilemma be replicated in other dominance-solvable games, such as 
the p-beauty contest or the centipede game? The level-$k$ and cognitive hierarchy literature \citep{nagel1995unraveling, camerer2004cognitive} provides natural tools for examining whether timing shifts the distribution of strategic types.

As a final remark, virtual observability can also be interpreted as a specific form of behavioral bias, suggesting that individuals may behave differently across extensive games that share the same reduced normal form. This contrasts with the standard game-theoretic perspective, in which reduced-normal-form invariance is regarded as a desirable property \citep{kohlberg1986strategic}. However, recent developments in behavioral game theory have shown that violations of reduced-normal-form invariance are a common feature of several behavioral solution concepts, including the agent quantal response equilibrium \citep{mckelvey1998quantal}, the dynamic cognitive hierarchy solution \citep{lin2022cognitive, linPalfrey2022DCH}, and the cursed sequential equilibrium \citep{fong2023cursed}. More broadly, our study of virtual observability contributes to the growing literature on such violations.




\bibliographystyle{ecta}

\bibliography{reference}

\newpage
\appendix

\renewcommand{\thesection}{Appendix \Alph{section}}

\setcounter{table}{0}
\renewcommand{\thetable}{A.\arabic{table}}

\section{Summary Statistics}
\label{sec:additional_tables}

\begin{table}[htbp!]
\centering
\caption{Summary Statistics of Claims in the Traveler's Dilemma}
\label{tab:summary_stat_claim}
\renewcommand{\arraystretch}{1.25}
\begin{tabular}{lccccc}
\hline
\multicolumn{1}{c}{Claims} &  & Obs & Mean & Median & S.D. \\ \hline
\multicolumn{6}{l}{\emph{A. Pooled Data}} \\
Simultaneous &  & 108 & 5.94 & 6.00 & 1.56 \\
Sequential &  &  &  &  &  \\
\multicolumn{1}{r}{First-Mover} &  & 54 & 5.57 & 5.25 & 1.55 \\
\multicolumn{1}{r}{\;\;\;\;Second-Mover} &  & 54 & 6.14 & 6.00 & 1.50 \\ \hline
\multicolumn{6}{l}{\emph{B. Sessions where Sequential played first}} \\
Simultaneous &  & 58 & 6.06 & 6.00 & 1.56 \\
Sequential &  &  &  &  &  \\
\multicolumn{1}{r}{First-Mover} &  & 29 & 5.59 & 5.00 & 1.63 \\
\multicolumn{1}{r}{Second-Mover} &  & 29 & 6.16 & 6.00 & 1.40 \\ \hline
\multicolumn{6}{l}{\emph{C. Sessions where Simultaneous played first}} \\
Simultaneous &  & 50 & 5.81 & 6.00 & 1.56 \\
Sequential &  &  &  &  &  \\
\multicolumn{1}{r}{First-Mover} &  & 25 & 5.56 & 5.50 & 1.49 \\
\multicolumn{1}{r}{Second-Mover} &  & 25 & 6.12 & 6.00 & 1.63 \\ \hline
\end{tabular}
\end{table}

\begin{table}[htbp!]
\centering
\caption{Summary Statistics of Investment and Return in 
the Trust Game}
\label{tab:my-table}
\renewcommand{\arraystretch}{1.25}
\begin{tabular}{rccccc}
\hline
\multicolumn{1}{c}{} &  & Obs & Mean & Median & S.D. \\ \hline
\multicolumn{6}{l}{\emph{Investment}} \\
\;\;\;\;\;\;Investor-First &  & 54 & 1.94 & 2.00 & 1.38 \\
Trustee-First &  & 54 & 1.98 & 2.00 & 1.22 \\ \hline
\multicolumn{6}{l}{\emph{Return}}\\
Investor-First &  & 54 & 5.52 & 6.00 & 2.23 \\
Trustee-First &  & 54 & 5.37 & 6.00 & 2.12 \\ \hline
\end{tabular}
\end{table}

\section{Omitted Statements and Proofs}
\label{sec:proof}

\begin{proposition}\label{prop:magical_td}
Under General Virtual Observability in the sequential Traveler's Dilemma, Traveler 1 will choose 8, and Traveler 2 will choose 7.5.
\end{proposition}

\noindent\textbf{Proof.} 
To derive the prediction of the General Virtual Observability in the sequential 
Traveler's Dilemma, we first derive Traveler 1's 
virtual conjecture about Traveler 2. For any $n_1 \in \{4, 4.5,...,8\}$, Traveler 
2's payoff from choosing $n_2$ is given by:
\begin{align*}
u_2(n_1, n_2) = 
\begin{cases}
n_2 + 1 & \text{if } n_2 < n_1, \\
n_1 & \text{if } n_2 = n_1, \\
n_1-1 & \text{if } n_2 > n_1. 
\end{cases}
\end{align*}
For any $n_1 \geq 4.5$, Traveler 2's best response 
is $n_1 - 0.5$, and for $n_1 = 4$, the best response is $4$. That is, 
Traveler 1's virtual conjecture is:
\begin{align*}
\tilde{\sigma}_2(\cdot | n_1) = 
\begin{cases}
\delta_{4} & \text{if } n_1 = 4, \\
\delta_{n_1 - 0.5} & \text{if } n_1 \geq 4.5,
\end{cases}
\end{align*}
where $\delta_{n_2}$ denotes the Dirac measure on Traveler 2 choosing $n_2$. 

Under General Virtual Observability, Traveler 1 best responds to $\tilde{\sigma}_2$, 
and therefore, Traveler 1's expected payoff from choosing $n_1$ is:
\begin{align*}
\mathbb{E}[\tilde{u}_1(n_1)] = 
\begin{cases}
4 & \text{if } n_1 = 4, \\
n_1  -1.5 & \text{if } n_1 \geq 4.5.
\end{cases}
\end{align*}
Consequently, Traveler 1's best response is $n_1 = 8$. Given this, 
Traveler 2's best response is $n_2 = 7.5$. This completes the proof.
$\blacksquare$

\bigskip

\begin{proposition}\label{prop:magical_trust}
Under General Virtual Observability in the investor-first sequential Trust Game, both the investor and the trustee choose 4, whereas in the trustee-first version, both choose 0.
\end{proposition}

\noindent\textbf{Proof.} 
In the \emph{investor-first} sequential Trust Game, for any $I \in 
\{0, 1, 2, 3, 4\}$, the trustee's payoff from choosing $R$ is given by:
\begin{align*}
u_t(I, R) = 
\begin{cases}
E+3I - 2R & \text{if } R \geq I, \\
E/2 & \text{if } R < I.
\end{cases}
\end{align*}
Therefore, for any $I \in \{0, 1, 2, 3, 4\}$, the trustee's best response is 
$R = I$. That is, the investor's virtual conjecture is 
$\tilde{\sigma}_t(\cdot | I) = \delta_I$. 
Under General Virtual Observability, the investor best responds to $\tilde{\sigma}_t$, 
and hence the investor's expected payoff from choosing $I$ is $E + I$. 
As a result, the optimal choice for the investor is $I = 4$, and the trustee 
correspondingly chooses $R = 4$.

In the \emph{trustee-first} sequential Trust Game, for any $R \in 
\{0, 1, 2, 3, 4\}$, the investor's payoff from choosing $I$ is given by:
\begin{align*}
u_i(I, R) = 
\begin{cases}
E - I + 2R & \text{if } I \leq R, \\
E/2 & \text{if } I > R.
\end{cases}
\end{align*}
Thus, for any $R \in \{0, 1, 2, 3, 4\}$, the investor's best response is 
$I = 0$. In other words, the trustee's virtual conjecture is 
$\tilde{\sigma}_i(\cdot \mid R) = \delta_0$. 
Under General Virtual Observability, the trustee best responds to $\tilde{\sigma}_i$, 
and hence the trustee's expected payoff from choosing $R$ is $-R$. 
As a result, the optimal choice for the trustee is $R = 0$, and the investor 
correspondingly chooses $I = 0$. This completes the proof. $\blacksquare$

\section{Experimental Instructions}
\label{sec:instruction}

This appendix provides the experimental instructions for the simultaneous and sequential 
Traveler's Dilemma, as well as the Trust Game. Since the instructions for the 
simultaneous and sequential Traveler's Dilemmas are nearly identical, we include only the
first and summary pages of the sequential games, the pages where the two versions differ.

\subsection{Traveler's Dilemma Instructions}

\noindent\textbf{Instructions for the Simultaneous Treatment, Page 1}

\begin{itemize}
    \item \textbf{Rounds:} The experiment consists of a single ``round,'' which will not be repeated.
    \item \textbf{Matching:} In this round, you will be matched with another person.
    \item \textbf{Interdependence:} The decisions that you and the other person make will 
    determine the amounts earned by each of you.
    \item \textbf{Choices:} You begin by choosing a number or \textbf{claim}. The person you are matched with will also choose their claim at the same time. You cannot see their claim while making yours, and vice versa.
    \item \textbf{Earnings:} The claims must be numbers, and the payoff to you will be the \textbf{minimum} of the two claims, plus an additional amount, a reward for the person making the lower claim and a penalty for the person making the higher claim, as explained below. To see some examples, press:
\end{itemize}

\bigskip
\noindent\textbf{Instructions for the Simultaneous Treatment, Page 2}

\begin{itemize}
    \item\textbf{Example 1:} If you choose a claim of 2 cents and the other also chooses 2, then the minimum is 2, and you each earn 2; there is no penalty or reward when the claims are equal.
    \item\textbf{Example 2:} Suppose that the penalty and reward amounts are each 1 cent. If one of you chooses 2 cents and the other chooses 3, then the minimum is 2 and the payoff is 2 plus the reward $= 2 + 1 = 3$ for the person who chose 2, and the payoff is 2 minus the penalty $= 2 -1 = 1$ for the person who chose 3.
    \item\textbf{Note:} The numbers used in the actual experiment to follow will be much larger than these numbers, which are for illustrative purposes only.
\end{itemize}

\bigskip
\noindent\textbf{Instructions for the Simultaneous Treatment, Page 3}
\bigskip

\noindent Now let's look at the actual numbers that determine earnings in the experiment.

\begin{itemize}
    \item \textbf{Allowable Claims:} At the beginning of each round, you will choose a claim that must be a number between and including \$4.00 and \$8.00 (in \$0.50 increments). The person who you are matched with will also choose a claim between and including \$4.00 and \$8.00 (in \$0.50 increments).
    \item \textbf{Earnings:} You will receive a monetary amount that equals the \textbf{minimum} of the two claims by you and the other person, plus a reward if your claim is lower, or minus a penalty if your claim is higher. If the claims are equal, there is no penalty or reward, and each person earns an amount that equals the common claim.
    \item \textbf{Penalty and Reward:} The penalty and reward amounts are equal to \$1.00. Thus the person with the lower claim receives this low claim plus \$1.00, and the person with the higher claim receives the low claim minus \$1.00.
\end{itemize}

\bigskip
\noindent\textbf{Instruction Summary for the Simultaneous Treatment, Page 4}

\begin{itemize}
    \item This process will only be played out once.
    \item You will be paired with another person, and you each may make your claim decision when you are ready, but you will not find out the other person's decision until both decisions have been made and confirmed.
    \item All ``claims'' must be between and including \textbf{\$4.00} and \textbf{\$8.00}. Each person earns an amount that equals the \textbf{minimum} of the two claims, plus a reward of \textbf{\$1.00} for the person with the \textbf{lower claim}, and minus a penalty of \textbf{\$1.00} for the person with the \textbf{higher claim}.
    \item There will be a \textbf{single round}, with no repetition.

\end{itemize}

\bigskip
\noindent\textbf{Instructions for the Sequential Treatment, Page 1}

\begin{itemize}
    \item \textbf{Rounds:} The experiment consists of a single ``round,'' which will not be repeated.
    \item \textbf{Matching:} In this round, you will be matched with another person.
    \item \textbf{Interdependence:} The decisions that you and the other person make will determine the amounts earned by each of you.
    \item \textbf{Choices:} You begin by choosing a number or \textbf{claim}. The person you are matched with will also choose a claim.
    \item \textbf{Decision Sequence:} One person in each matched pair has been designated to be a \textbf{first mover} who will choose a claim first. The other person is a \textbf{second mover} who must wait until the first person decides and confirms before choosing a claim. After the second mover has made and confirmed a decision, both decisions are revealed and earnings for the round are calculated.
    \item \textbf{Earnings:} The claims must be numbers, and the payoff to you will be the \textbf{minimum} of the two claims, plus an additional amount, a reward for the person making the lower claim and a penalty for the person making the higher claim, as explained below. To see some examples, press:
\end{itemize}

\bigskip
\noindent\textbf{Instruction Summary for the Sequential Treatment, Page 4}

\begin{itemize}
    \item This process will only be played out once.
    \item One person has in each pair has been designated to be a \textbf{first mover}, and the other will be a \textbf{second mover}. The first mover must make and confirm a claim decision before the second mover makes their decision, but the second mover does not see the first mover's decision before choosing their own claim decision.
    \item You will be a ***(first mover or second mover)*** in all rounds.
    \item All ``claims'' must be between and including \textbf{\$4.00} and \textbf{\$8.00}. Each person earns an amount that equals the \textbf{minimum} of the two claims, plus a reward of \textbf{\$1.00} for the person with the \textbf{lower claim}, and minus a penalty of \textbf{\$1.00} for the person with the \textbf{higher claim}.
    \item There will be a \textbf{single round}, with no repetition.

\end{itemize}

\subsection{Trust Game Instructions}

\noindent\textbf{Instructions, Page 1}

\begin{itemize}
    \item \textbf{Single Decision:} The experiment consists of a single \textbf{round} in which you are matched with another person. Note: You will only make a single decision in this experiment.
    \item \textbf{Interdependence:} The decisions that you and the other person make will determine the amounts earned by each of you (details to follow).
    \item \textbf{Pass/Keep Decisions:} One of you will receive an endowment of money, \textbf{\$4.00} and will decide how much (if any) to pass on to the other person and how much (if any) to keep. All money passed gets multiplied by 3 before it is delivered to the \textbf{receiver}, who also receives an endowment of \textbf{\$4.00}.
    \item \textbf{Return Decisions:} Before learning of the exact amount passed, the receiver must decide how much money to pass back (if any) to the \textbf{passer}. Therefore, neither person knows the other's decision at the time that they make their own decisions. These pass/keep decisions determine earnings for the round, as explained next.
\end{itemize}

\bigskip

\noindent\textbf{Instructions, Page 2}

\begin{itemize}
    \item \textbf{Role:} You have been randomly assigned to be a \textbf{Passer} or \textbf{Receiver}. The passer will begin each round with an amount of money, \textbf{\$4.00} and will decide how much to keep and how much to pass (\$0, \$1, \$2, \$3, or \$4) to the other person (\textbf{Receiver}). All money that is passed to the receiver will be multiplied by \textbf{3}. The receiver must decide how much of the tripled amount to send back, in multiples of \$2: (\$0, \$2, \$4, \$6, or \$8), before seeing the amount that was sent by the passer. So neither person will know the other's decision before you make their own decision.

    \item \textbf{Order of Decisions:} Even though neither person will be able to observe the other's pass or return decision, one of you will be designated to make their decision first, and the other will be permitted to make their decision only after the first mover has recorded and confirmed a decision, but before this earlier decision is revealed. You will be told at the beginning of the round whether you are the first mover or the second mover for the round. The first mover could be designated to be either the passer or the receiver.

    \item \textbf{Earnings from Pass/Keep Process:}
    \begin{itemize}
        \item[1)] If the return amount is less than 2 times the pass amount, then the pass is invalidated, in which case the passer earns a reduced endowment of \$2.00 and the receiver earns a reduced endowment of \$2.00.
        \item[2)] If the return amount is greater than or equal to 2 times the pass amount, the pass is made, and the passer earns the amount kept initially plus the amount that is passed back by the receiver. All money sent by the passer is multiplied by \textbf{3} before being delivered, and the receiver earns the amount kept at this stage plus their full endowment of \$4.00.
    \end{itemize}
\end{itemize}

\bigskip

\noindent\textbf{Instructions, Page 3}

\begin{itemize}
    \item \textbf{Examples:} If the passer sends \$0, any amount returned is valid (since return $\geq 2$ times pass amount). If the passer sends \$1, the receiver must return at least \$2 for the pass to be valid (for return $\geq 2$ times pass amount). If the passer sends \$3, the receiver must return at least \$6 for the pass to be valid, etc.

    \item \textbf{Another Example:} Suppose passer sends \$3 and receiver sends \$8. Earnings for the passer equal \$9 (the \$1 amount kept ($\$4-\$3$) plus the \$8 amount sent by the receiver. Earnings for the receiver equal \$5 (the amount kept $\$3 \times 3 - \$8 $, plus the endowment \$4). Suppose the sender sends \$3 and the receiver sends \$4. The pass is invalidated since \$4 $<$ 2 times the \$3 pass, and therefore both passer and receiver earn reduced endowment amounts of \$2.
\end{itemize}

\bigskip
\noindent\textbf{Instruction Summary, Page 4}

\begin{itemize}
    \item \textbf{One Decision Only:} Please remember that you are matched with another participant, and that this process will \textbf{not} be repeated, you make only \textbf{one decision}.

    \item \textbf{Decisions:} The passer begins each round with an endowment of \textbf{\$4.00} and must decide how much (if any) to keep and how much (if any) to pass to the receiver. What is passed gets \textbf{tripled} before being delivered. The receiver in each pair decides how much (if any) to keep and how much (if any) to pass back. One person in each pair will be designated to make their decision (pass or return) before the other. But each person makes their own decision without having observed the other's decision.

    \item\textbf{Earnings with Sufficient Return:} If the return amount is \textbf{greater than or equal to 2 times the pass amount}, then the passer earns the amount kept initially from their endowment plus the amount passed back, and the receiver earns the amount kept plus their endowment.

    \item\textbf{Earnings with Insufficient Return:} If the return amount is \textbf{less than 2 times the pass amount}, the pass is invalidated, and each person only earns half of their endowment: \$2.00 for the passer and \$2.00 for the receiver. 

    \item\textbf{Finish:} After you and the person you are matched with have made your decisions, earnings will be determined but will not be announced until the administrator releases earnings information after other tasks have been completed.
\end{itemize}

\end{document}